\documentclass{PoS}

\usepackage{amsfonts}
\usepackage{amsmath}
\usepackage{subfigure} 

\graphicspath{{./figs/}}

\title{
$B_K$ from improved staggered fermions using SU(3) chiral
perturbation theory
}

\ShortTitle{
  SU(3) ChPT analysis of $B_K$
}

\author{\speaker{Kwangwoo Kim}, Yong-Chull Jang,
  Hyung-Jin Kim, Jangho Kim, Boram Yoon, Weonjong Lee\\
  Lattice Gauge Theory Research Center, CTP, and FPRD, \\
  Department of Physics and Astronomy,
  Seoul National University, Seoul, 151-747, South Korea \\
  E-mail: \email{wlee@snu.ac.kr}}

\author{Taegil Bae \\
  Korea Institute of Science and Technology Information,
  Daejeon, 305-806, South Korea \\
  E-mail: \email{esrevinu@gmail.com}}

\author{Chulwoo Jung \\
  Physics Department, Brookhaven National Laboratory,
  Upton, NY11973, USA \\
  E-mail: \email{chulwoo@bnl.gov}}

\author{Jongjeong Kim \\
  Physics Department,
  University of Arizona,
  Tucson, AZ 85721, USA \\
  E-mail: \email{rvanguard@gmail.com}}

\author{Stephen R. Sharpe\\
  Physics Department, University of Washington, Seattle, WA 98195-1560 \\
  E-mail: \email{sharpe@phys.washington.edu}}

\author{SWME Collaboration}

\abstract{ 
  We present recent progress in our calculation of
  $B_K$ with improved staggered fermions using chiral extrapolations
  based on SU(3) staggered chiral perturbation theory.
  We have accumulated significantly higher statistics on the coarse,
  fine, and ultrafine MILC asqtad lattices.  
  This leads to a reduction in statistical error and an improved
  continuum extrapolation.
  Our updated result is $\hat{B}_K = B_K(\text{RGI}) = 0.737 \pm
  0.003(\text{stat}) \pm 0.046 (\text{sys})$.
  This is consistent with the result obtained using chiral extrapolations
  based on SU(2) staggered chiral perturbation theory, although the
  total error is somewhat larger with the SU(3) analysis.
}

\FullConference{
  The XXIX International Symposium on Lattice Field Theory - Lattice 2011\\
  July 10-16, 2011\\
  Squaw Valley, Lake Tahoe, California
}

\begin{document}

\section{Introduction} 
Calculations of the kaon mixing parameter $B_K$ have reached
a mature stage with several results available having all errors
controlled and small. Our calculation uses improved staggered 
fermions---HYP-smeared valence on asqtad sea---and at present
achieves total error of $\sim5\%$~\cite{BKPRD,Lat2011su2}.
Our main analysis uses chiral fitting functions derived from
SU(2) staggered chiral perturbation theory (SChPT),
and is updated in Ref.~\cite{Lat2011su2}.
As a check on this result, we also carry out an analysis using
fit forms from SU(3) SChPT, and in the present report we update
the results from this analysis.
In particular, we focus on the progress since our publication
\cite{BKPRD} and last year's lattice proceedings~\cite{ref:wlee-2010-2}.

In Table~\ref{tab:milc-lat} we show the current set of ensembles
on which we have done the SU(3) analysis, together with the
resulting values for $B_K$.
In the last year,
we have accumulated higher statistics on the C2, C5, F1, F2, and U1
ensembles.
The most important improvements are those for the F1 and U1 ensembles,
since these are used for continuum extrapolation.
As described in Ref.~\cite{Lat2011su2},
the 9-fold increase in statistics on the F1 ensemble significantly
impacts the continuum extrapolation in the SU(2) analysis.
It forces us to fit to only the three smallest
lattice spacings, excluding the coarse ensembles.
Our main aim here is to show how the increase in statistics
impacts the SU(3) analysis.

\begin{table}[h!]
\begin{center}
\begin{tabular}{c | c | c | c | c | c | c}
\hline
$a$ (fm) & $am_\ell/am_s$ & geometry & ID & ens $\times$ meas 
& $B_K$ (N-BB1) & $B_K$ (N-BB2) \\
\hline
0.12 & 0.03/0.05  & $20^3 \times 64$ & C1 & $564 \times 1$ &  0.555(12) & 0.564(17) \\
0.12 & 0.02/0.05  & $20^3 \times 64$ & C2${}^*$ & $486 \times 9$ &  0.538(12) & 0.535(17) \\
0.12 & 0.01/0.05  & $20^3 \times 64$ & C3 & $671 \times 9$ &  0.562(6)  & 0.592(14) \\
0.12 & 0.01/0.05  & $28^3 \times 64$ & C3-2 & $275 \times 8$ &  0.575(6)  & 0.595(13) \\
0.12 & 0.007/0.05 & $20^3 \times 64$ & C4 & $651 \times 10$ & 0.564(5)  & 0.598(13) \\
0.12 & 0.005/0.05 & $24^3 \times 64$ & C5${}^*$ & $509 \times 9$ &  0.576(5) & 0.598(12) \\
\hline
0.09 & 0.0062/0.031 & $28^3 \times 96$ & F1${}^*$ & $995 \times 9$ & 0.536(3) & 0.561(10) \\
0.09 & 0.0031/0.031 & $40^3 \times 96$ & F2${}^*$ & $850 \times 1$ & 0.539(7) & 0.544(13)\\
\hline
0.06 & 0.0036/0.018 & $48^3 \times 144$ & S1 & $744 \times 2$ & 0.535(6) & 0.560(11)\\
0.06 & 0.0025/0.018 & $56^3 \times 144$ & S2${}^{\#}$ & $198 \times 9$ & -NA- & -NA-\\
\hline
0.045 & 0.0028/0.014 & $64^3 \times 192$ & U1${}^*$ & $705 \times 1$ & 0.543(4) & 0.554(8)\\
\hline
\end{tabular}
\end{center}
\caption{MILC asqtad ensembles used in the calculation.  Ensembles
marked with a ${}^*$ have improved statistics compared to last year,
while those marked with a ${}^\#$ are new. Results for $B_K(\mu=2\
{\rm GeV})$ using both N-BB1 and N-BB2 fits are given. See the text for
discussion of these fits.}
\label{tab:milc-lat}
\end{table}

\section{Fitting and Results}

In our numerical study, our lattice kaon is composed of valence
(anti-)quarks with masses $m_x$ and $m_y$.
On each MILC asqtad ensemble, we use 10 valence masses:
\begin{equation}
am_x,\ am_y = am_s \times n/10 
\qquad \text{with} \qquad
n = 1,2,3,\ldots,10
\end{equation} 
where $am_s$ is the nominal 
strange sea quark mass of the given lattice ensemble.
Hence, we have 55 combinations of the lattice kaon: 10 degenerate
($m_x = m_y$) and 45 non-degenerate ($m_x \ne m_y$).

An important difference between the SU(2) and SU(3) analyses 
is the number of the mass combinations that are included.
In the SU(2) analysis, we use only those
in which the valence $d$ quark is much lighter than the valence $s$ quark.
Specifically, we use the lightest 4 or 5 values of $m_x$
and the heaviest 3 values of $m_y$.
In the SU(3) analysis, by contrast, we use all 55 combinations.
While this makes better use of our data, it does so at 
considerable cost. First, quite a few of our mass combinations are
in the regime where next-to-leading order (NLO)
SU(3) ChPT is beginning to break down.
Second, the fit forms in SU(3) 
SChPT contain very many fit parameters~\cite{VdWSS}
and have to be simplified by hand in order to be practical.
In the end, as explained in Ref.~\cite{BKPRD}, we came up with
two different schemes for fitting, both using Bayesian constraints
on parameters which arise due to discretization errors, but
doing so in somewhat different ways. We focus here on the
results for these two schemes, which we call ``N-BB1'' and ``N-BB2''.
Both schemes are based on NLO SChPT with the addition of
a single analytic next-to-next-to-leading order term.

In the N-BB1 scheme, we fit the data in two stages.
First, we fit the 10 degenerate mass combinations to the functional
form of Eq.~(62) of Ref.~\cite{BKPRD} using the Bayesian method.
We constrain the coefficient of the term arising from
discretization and matching errors assuming that discretization
errors dominate, so that
$c_4 \propto ( a \Lambda_\textrm{QCD} )^2$.
Second, we fit all 55 combinations
to the fitting functional form of Eq.~(63) of Ref.~\cite{BKPRD}
using the Bayesian method.
In this second stage, we make similar assumptions concerning the
size of terms arising from discretization errors, and also
input the results from the degenerate fit (for a subset of the
total set of parameters) as Bayesian constraints.

The N-BB2 scheme differs only in that we assume that
matching errors dominate in the low-energy coefficients arising
from discretization and matching errors, so that, for example,
$c_4 \propto \alpha_s^2$, 
(where we evaluate $\alpha_s$ at the scale of $\mu=1/a$ in
the $\overline{\text{MS}}$ scheme).  
%
%

\begin{figure}[htb!]
\centering
\subfigure[Lattice 2010]{\includegraphics[width=0.49\textwidth]
  {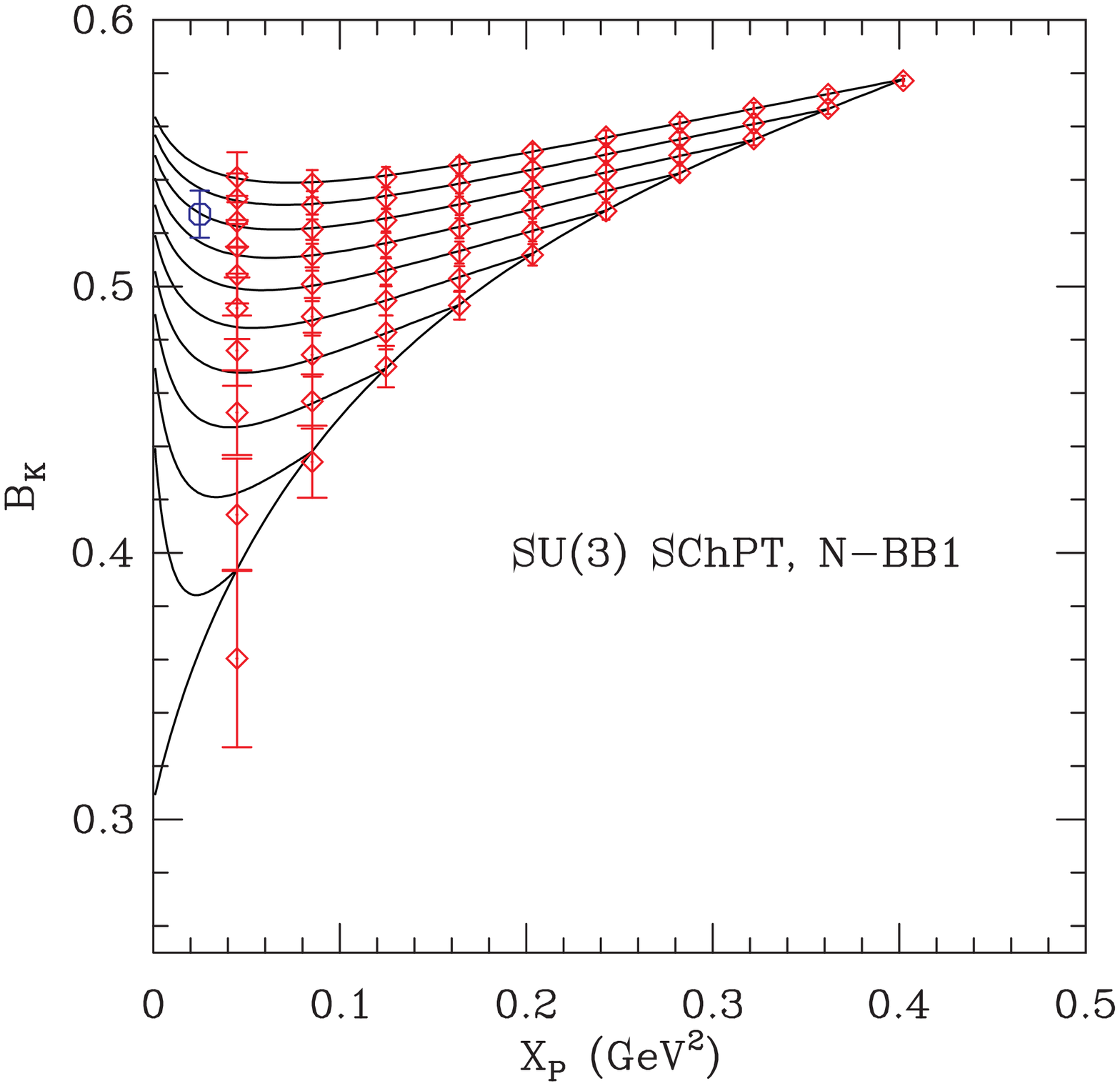}}
\subfigure[Lattice 2011]{\includegraphics[width=0.49\textwidth]
  {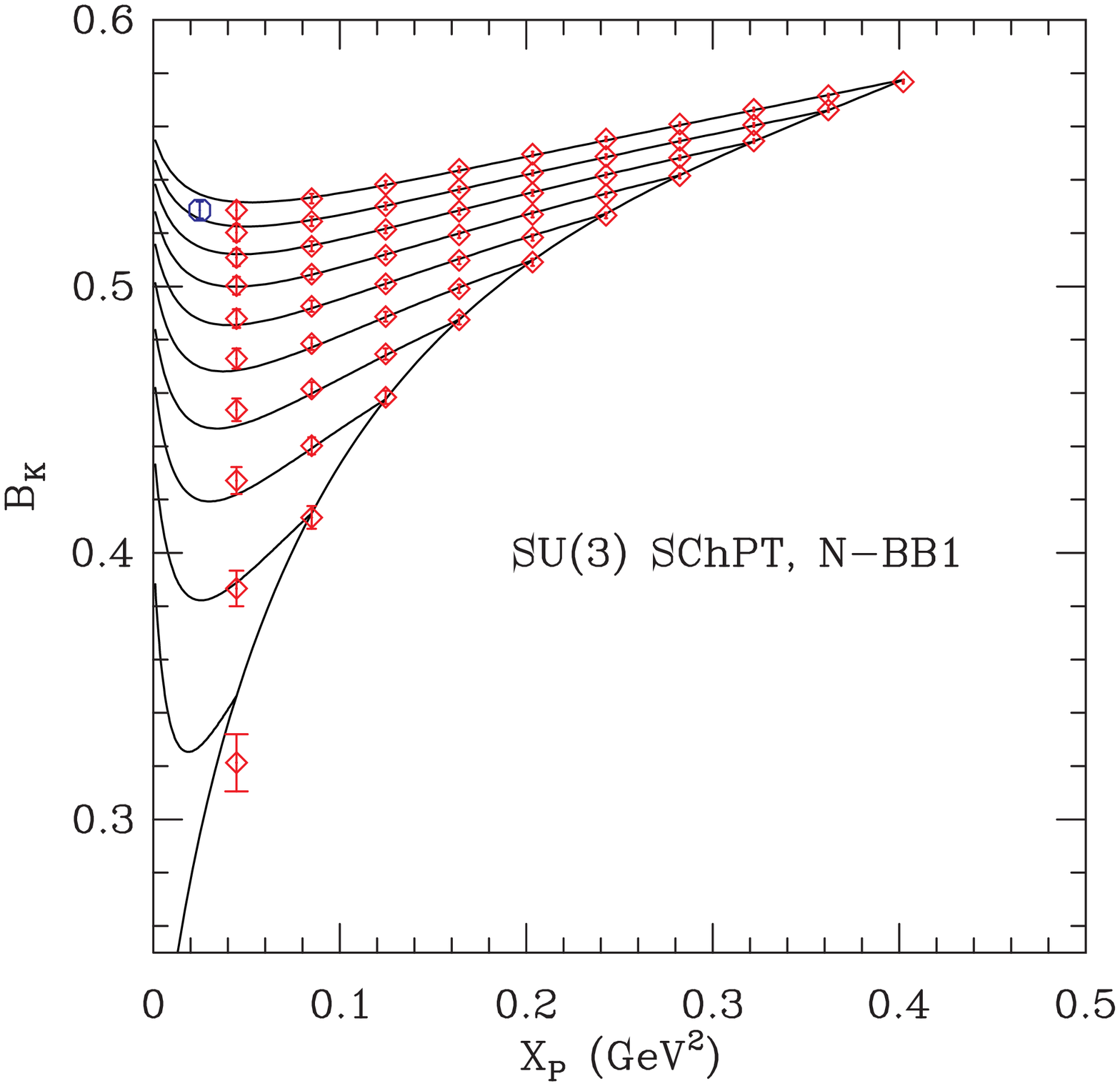}}
\caption{$B_K(1/a)$ versus $X_P$ (squared mass of pion composed of valence
  $x$ and $\bar x$) for the F1 ensemble, using N-BB1 fits.
  The left panel uses 1 measurement/configuration, while the
  right panel uses 9 measurements/configuration.
  Red diamonds show the data, while the blue octagon shows the
  result obtained after extrapolation to physical quark masses
  with all taste-breaking lattice artifacts removed.}
\label{fig:su3,N-BB1,F1}
\end{figure}
In Fig.~\ref{fig:su3,N-BB1,F1}, we show how the 9-fold increase
in statistics impacts the N-BB1 fits on the F1 ensemble.
As expected, errors have been reduced by a factor of $\sim 3$,
and this holds also for the value for $B_K$ 
that results after extrapolation
to physical valence-quark masses and removal of taste-breaking
lattice artifacts.
This final value (shown as a blue octagon in the figure) has
shifted up by about $0.2\sigma$.

The changes in the N-BB2 fits (not shown) are different:
the final value is shifted up by $1.8\sigma$ while
the error is reduced only by a factor of $\sim 1.2$.

\section{Continuum extrapolation}
\begin{figure}[htb!]
\centering
\subfigure[N-BB1]{\includegraphics[width=0.49\textwidth]{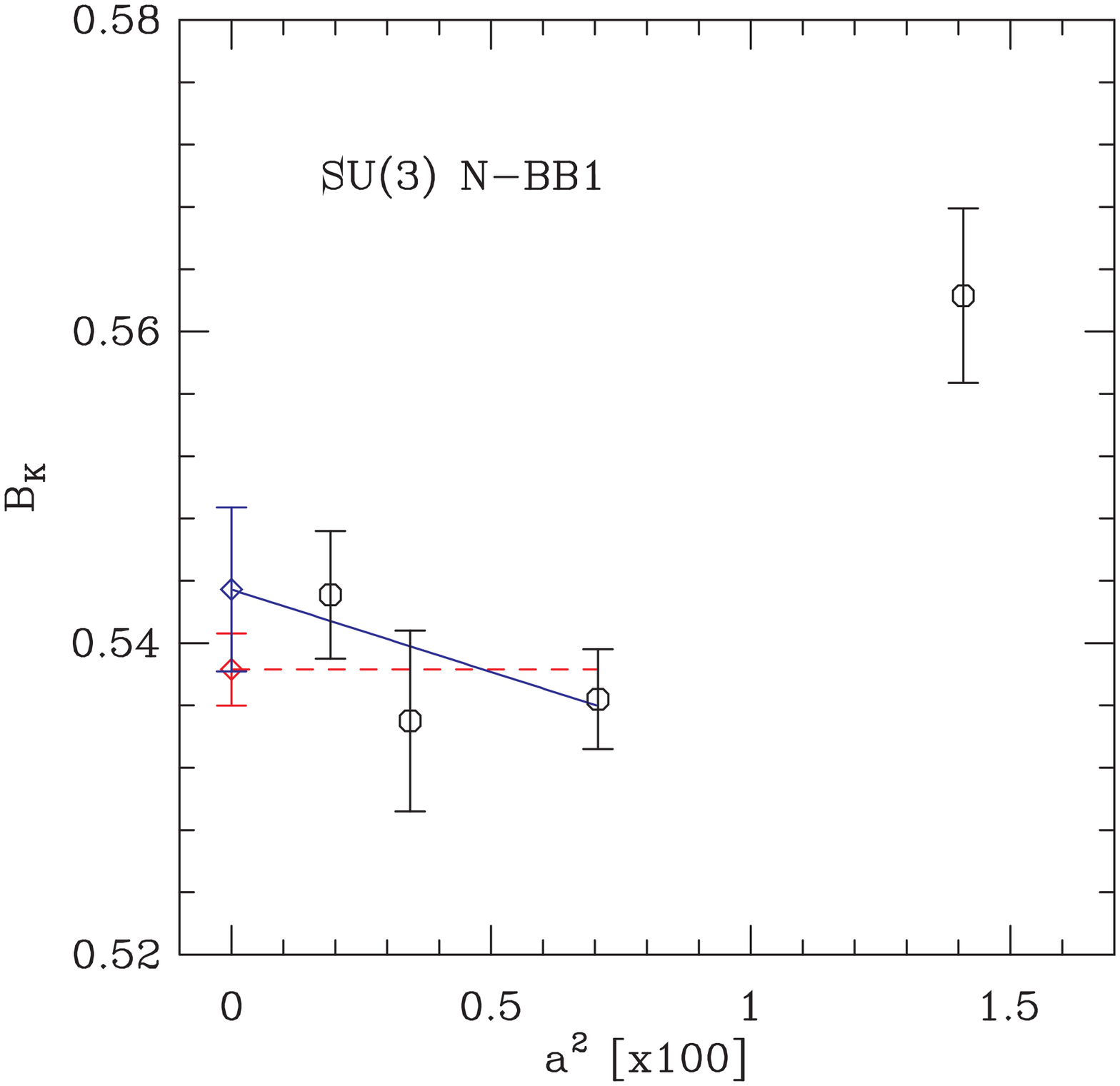}}
\subfigure[N-BB2]{\includegraphics[width=0.49\textwidth]{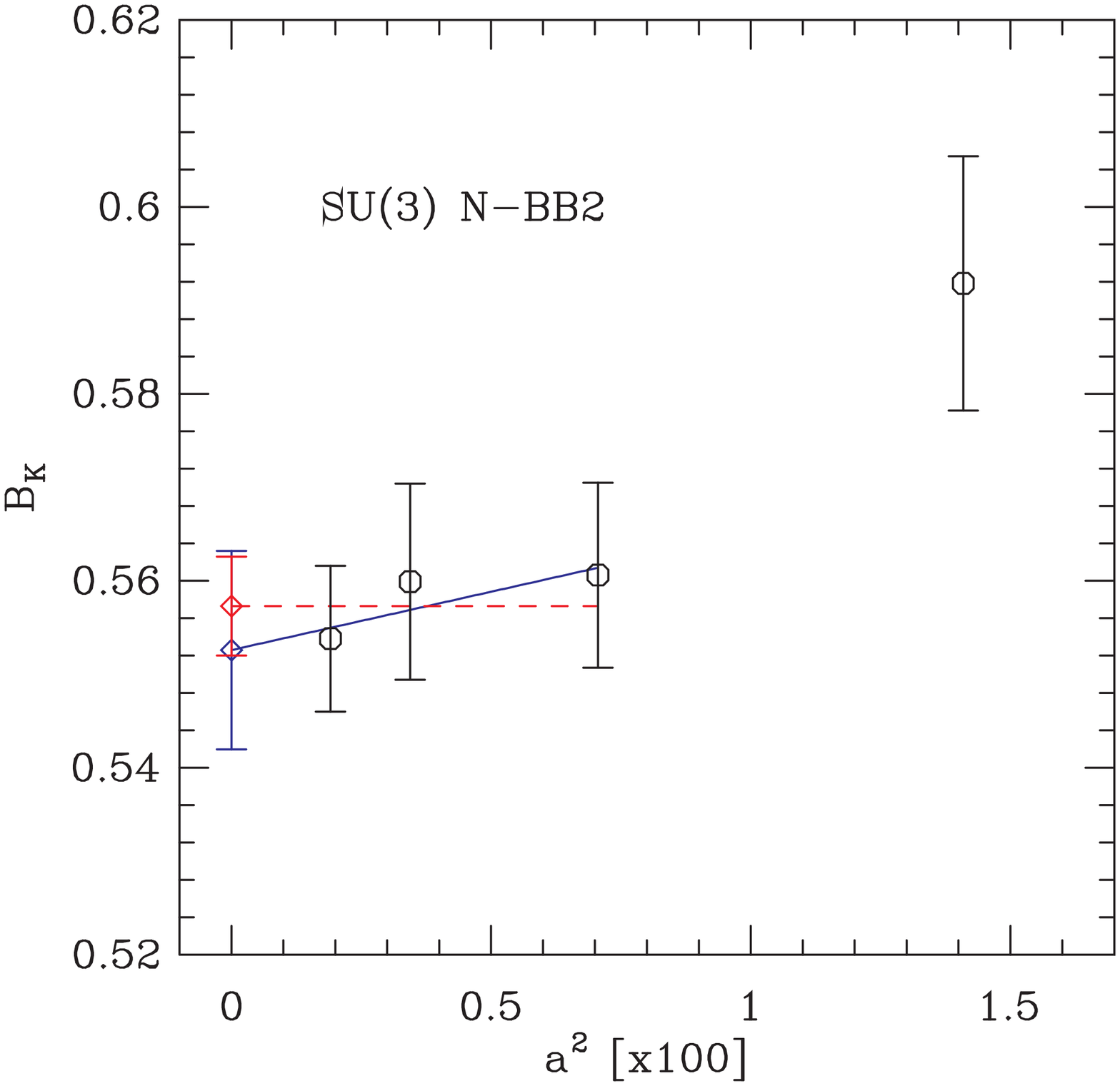}}
\caption{ $B_K(\text{NDR}, \mu=2\text{ GeV})$ as a function of $a^2$ 
(in fm $\times 100$) for the N-BB1 fit (left) and the N-BB2 fit (right), 
showing constant and linear extrapolations to the smallest three
values of $a$.
}
\label{fig:bk:a^2:bayes}
\end{figure}
In Fig~\ref{fig:bk:a^2:bayes}, we present our updated
results for the continuum extrapolation using
both N-BB1 and N-BB2 fits.
In the case of the N-BB1 fit, it is not possible to obtain a good
fit to all four values of $a$ using 
a simple fitting functional form, e.g. $c_1+c_2 a^2+ c_3 a^4$,
with physically reasonable values for the coefficients.
This is the same issue that arises in the SU(2) fits, and is
discussed in that case in more detail in the companion 
proceedings~\cite{Lat2011su2}.
We proceed by fitting only to the smallest three values of $a$,
using either constant or linear fits.

For the N-BB2 fit we do obtain reasonable fits to
all four values of $a$, but, in order to compare with the N-BB1 fits,
we also use only the smallest three values of $a$.
\begin{table}[bht!]
\begin{center}
\begin{tabular}{c | c | c }
\hline
fit type & constant fit  &  linear fit \\
\hline
N-BB1 & 0.5383(23) & 0.5344(53) \\
N-BB2 & 0.5573(53) & 0.5526(106)  \\
\hline
\end{tabular}
\end{center}
\caption{Results for $B_K(\mu=2\;{\rm GeV}$ 
to data from N-BB1 and N-BB2 chiral fits using both constant and
linear continuum extrapolations (in $a^2$).}
\label{tab:bk:n-bb1,n-bb2}
\end{table}

In Table~\ref{tab:bk:n-bb1,n-bb2}, we summarize the results of these
continuum extrapolations.
We use the constant fit to the N-BB1 data for our central value
and the difference between this and the result from the corresponding
fit to the N-BB2 data as an estimate of the fitting systematic.
Note that the difference between N-BB1 and N-BB2 fits is much
larger than the difference between constant and linear extrapolations.
This reflects the uncertainty in SU(3) fitting caused by
the large number of parameters related to discretization
and matching errors.

%
%
%
%
%
%
%

\begin{figure}[tbh!]
\centering
\subfigure[SU(3) analysis]{\includegraphics[width=0.49\textwidth]
  {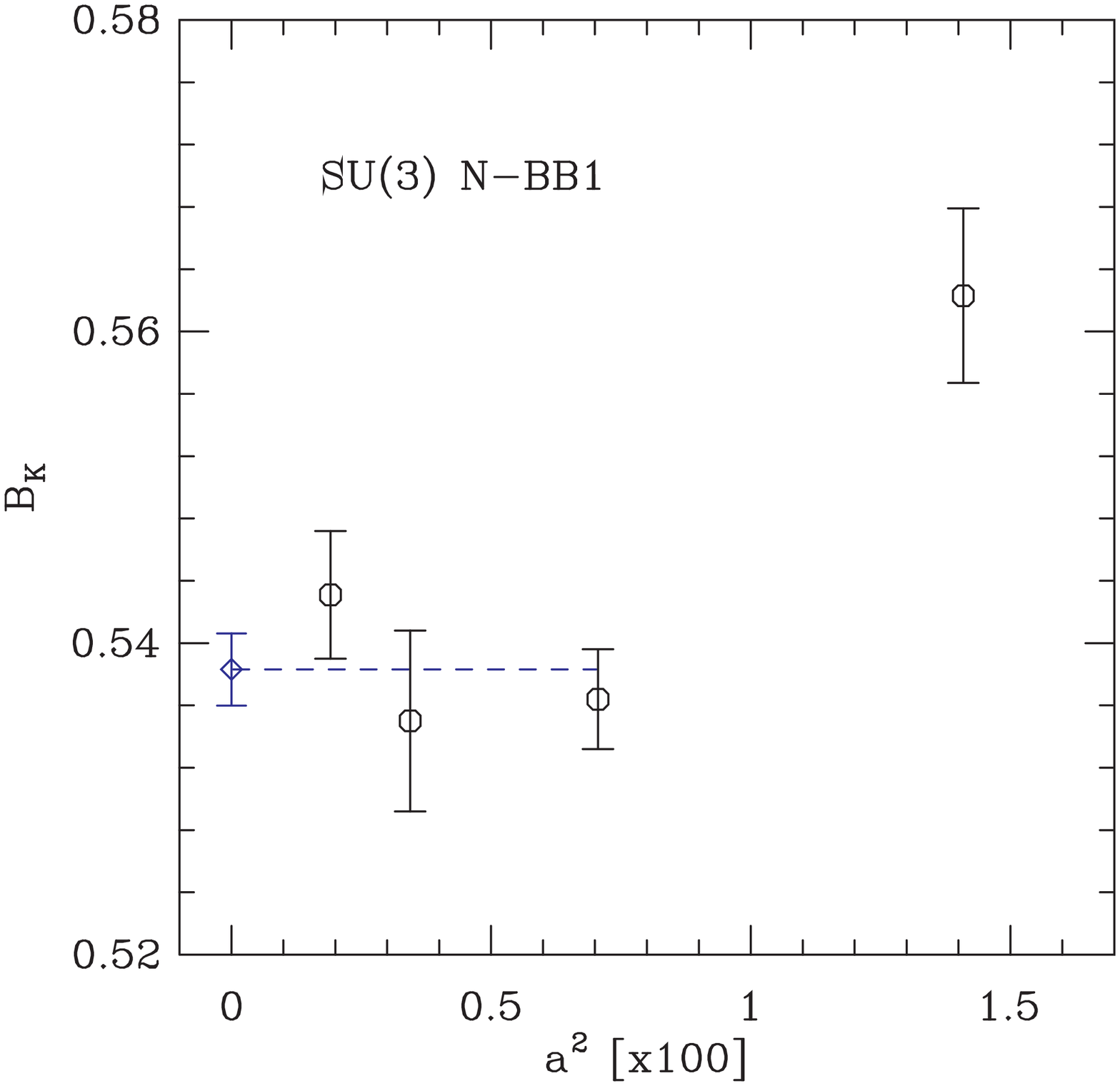}}
\subfigure[SU(2) analysis]{\includegraphics[width=0.49\textwidth]
  {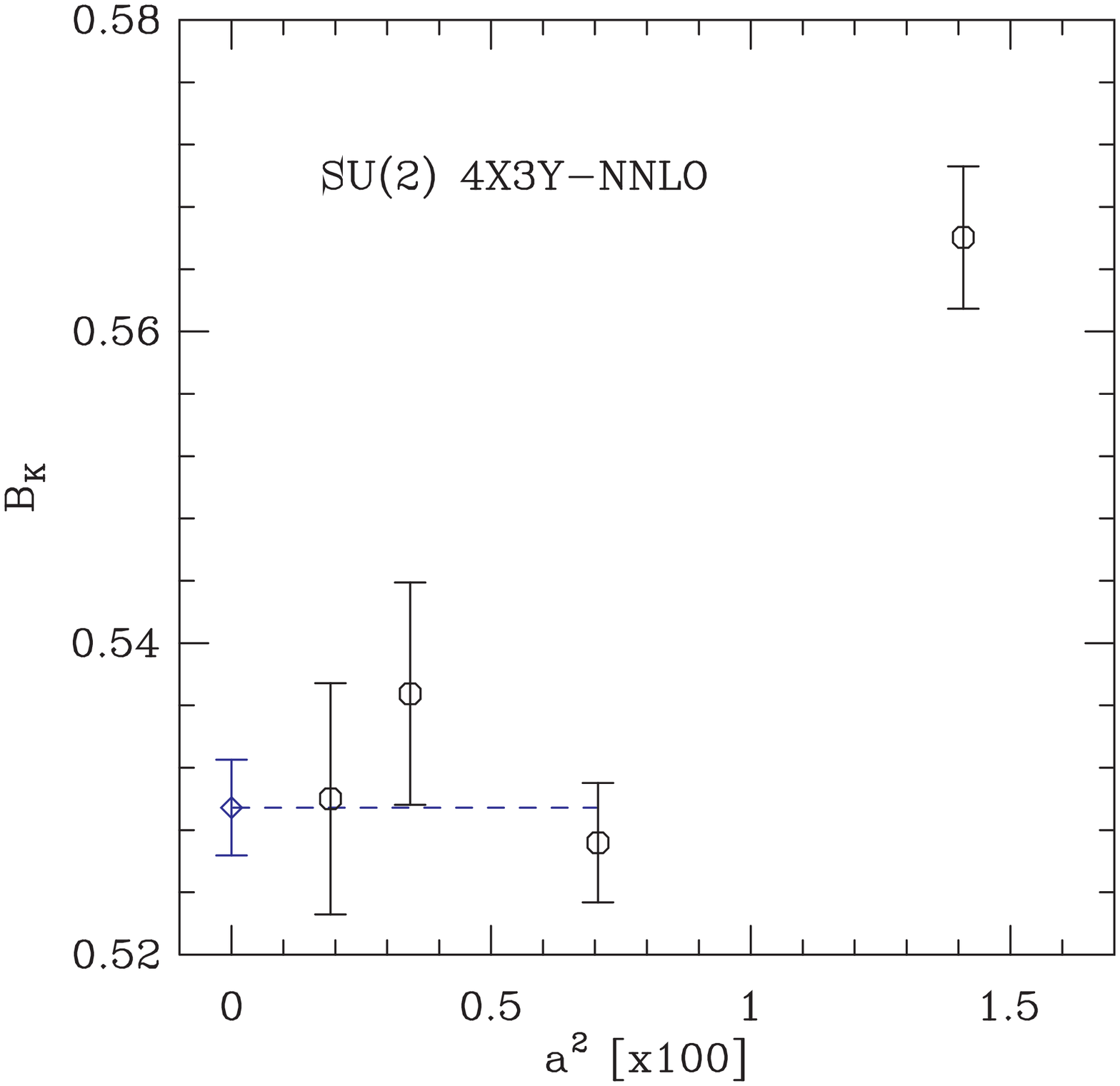}}
\caption{ $B_K(\text{NDR}, \mu=2\text{ GeV})$ as a function of $a^2$
  (in fm $\times 100$) from the SU(3) analysis (N-BB1 fit) in the left panel
and from the SU(2) analysis in the right panel. 
Constant fits to the three smallest values of $a=0$ are shown.  }
\label{fig:bk:a^2:su3-su2}
\end{figure}

%
%
%
In Fig.~\ref{fig:bk:a^2:su3-su2} we compare the results from
the N-BB1 fits to those using our preferred SU(2) fitting approach.
There is reasonable consistency point by point.

\section{Error Budget and Conclusions}
%
%
%

\begin{table}[tbh!]
\centering
\begin{tabular}{ l | l l r }
\hline \hline
cause & error (\%) & memo  & status \\
\hline
statistics       & 0.43  & N-BB1 fit & update \\
matching factor  & 4.4   & $\Delta B_K^{(2)}$ (U1) & update \\
discretization   & 0.95  & diff.~of const and linear extrap & update\\
fitting (1)      & 0.36  & diff.~of N-BB1 and N-B1 (C3) & \cite{BKPRD}\\
fitting (2)      & 3.5   & diff.~of N-BB1 and N-BB2 at $a=0$ & update \\
$a m_l$ extrap   & 1.0   & diff.~of (C3) and linear extrap & \cite{BKPRD}\\
$a m_s$ extrap   & 0.5   & constant vs. linear extrap & \cite{BKPRD}\\
finite volume    & 2.3   & diff.~of $20^3$ (C3) and $28^3$ (C3-2)& \cite{BKPRD}\\
$r_1$            & 0.12  & error propagation from $r_1$ & \cite{BKPRD}\\
\hline \hline
\end{tabular}
\caption{Error budget for $B_K$ obtained using SU(3) SChPT fitting.
  \label{tab:su3-err-budget}}
\end{table}
In Table \ref{tab:su3-err-budget}, we list various sources of
error in the SU(3)-based calculation of $B_K$.
Many of the smaller errors are unchanged from Ref.~\cite{BKPRD},
and we refer to that paper for an explanation of how they
are estimated.

The major changes to the errors are as follows.
The statistical error has been reduced from 1.4\% in Ref.~\cite{BKPRD}
to 0.4\%, due to the use of more measurements.
The ``matching'' error---due to our use of one-loop matching
between lattice and continuum operators---has been reduced from
5.5\% in Ref.~\cite{BKPRD} to 4.4\%. This is simply due to the
addition of the smallest lattice spacing, for our estimate
of the percentage error is $\alpha_s(\mu=1/a_{\rm min})^2$.
The discretization error has also been reduced (from 2.2\%), 
since our smallest value of $a$ is closer to the continuum.
Finally, the ``fitting (2)'' error---that due to the
uncertainty in the size of SChPT coefficients introduced by
discretization and matching errors---has changed from 5.3\%
to 3.5\%. Nevertheless, this error remains large, and along
with the matching error, dominates the total error.
This large fitting systematic is a reflection of the difficulties
in using SU(3) SChPT and is reason why we think that the SU(2)
analysis is more reliable.

Our present value from the SU(3) analysis is
\begin{equation}
\begin{array}{l l}
  B_K(\text{NDR}, \mu = 2 \text{ GeV}) & = 0.5383 \pm 0.0023 \pm 0.0337\,,
  \qquad {\rm SU(3)\ fit}
\\
  \hat{B}_K = B_K(\text{RGI}) & = 0.7371 \pm 0.0032 \pm 0.0461\,,
  \qquad {\rm SU(3)\ fit}
\end{array}
\end{equation}
where the first error is statistical and the second systematic.
The total error is 6.3\%.
This should be compared to our updated SU(2) result~\cite{Lat2011su2}
\begin{equation}
\hat{B}_K = B_K(\text{RGI}) = 
0.725 \pm 0.004(\text{stat}) \pm 0.038(\text{sys}) 
\qquad {\rm SU(2) \ fit}\,.
\end{equation}
We see that, although the SU(3) result has a smaller statistical
error, it has a significantly larger systematic error.
The most important observation, however, is that the two
results are consistent.


\section{Acknowledgments}
C.~Jung is supported by the US DOE under contract DE-AC02-98CH10886.
The research of W.~Lee is supported by the Creative Research
Initiatives Program (3348-20090015) of the NRF grant funded by the
Korean government (MEST). 
W.~Lee would like to acknowledge the support from KISTI supercomputing
center through the strategic support program for the supercomputing
application research [No. KSC-2011-C3-03].
The work of S.~Sharpe is supported in part by the US DOE grant
no.~DE-FG02-96ER40956.
Computations were carried out in part on QCDOC computing facilities of
the USQCD Collaboration at Brookhaven National Lab, and on the DAVID
GPU clusters at Seoul National University. The USQCD Collaboration are
funded by the Office of Science of the U.S. Department of Energy.

\end{document}